\documentclass[oldversion]{aa}
\usepackage{graphicx}
\usepackage{txfonts}
\usepackage{natbib}
\usepackage{url}

\begin{document}
   \title{Departures from LTE for neutral Li in late-type stars\thanks{Tables 2 and 3 are only available in electronic form
at the CDS via anonymous ftp to cdsarc.u-strasbg.fr (130.79.128.5) or via http://cdsweb.u-strasbg.fr/.}}

   \author{K. Lind\inst{1}\and
           M. Asplund\inst{2}\and
           P.\,S.\ Barklem\inst{3}}

   \institute{European Southern Observatory (ESO), Karl-Schwarzschild-Strasse 2,
              857 48 Garching bei M\"unchen, Germany\\
              \email{klind@eso.org}\and
              Max-Planck-Institut f\"ur Astrophysik, Karl-Schwarzschild-Strasse 1,
              857 41 Garching bei M\"unchen, Germany\and
              Department of Physics \& Astronomy, Uppsala University, Box 515, 751 20 Uppsala, Sweden
             }
   \date{Received 2009 March 28; accepted 2009 May 17}

\abstract{
We perform non-LTE calculations of lithium in late-type stars for a wide range of stellar parameters, including quantum mechanical cross-sections for collisions with neutral hydrogen and the negative hydrogen ion. Non-LTE abundance corrections for the lithium resonance line at 670.7\,nm and the subordinate line at 610.3\,nm, are calculated using 1D MARCS model atmospheres spanning a grid $T_{\rm eff}=[4000,8000]$\,K, $\log{g}=[1.0,5.0]$, and $\rm[Fe/H]=[0.0,-3.0]$, for lithium abundances in the range $A\rm(Li)=[-0.3,4.2]$. The competing effects of ultraviolet over-ionization and photon losses in the resonance line govern the behaviour of the non-LTE effects with stellar parameters and lithium abundance. The size and sign of the non-LTE abundance corrections vary significantly over the grid for the 670.7\,nm line, but are typically positive and below 0.15\,dex for the 610.3\,nm, line. The new collisional data play a significant role in determining the abundance corrections.}

   \keywords{Stars: abundances -- Stars: late-type -- Line: formation 
            }

   \maketitle

\section{Introduction}
Stellar lithium abundances are continuing to attract strong interest among the astronomical community. In particular, many studies are devoted to constraining and finding explanations for the famous ``Spite plateau'' of warm metal-poor halo stars, first discovered by \citet{Spite82}. For recent studies of lithium in halo field stars see e.g. \citet{Ryan01}, \citet{Charbonnel05a}, \citet{Asplund06}, \citet{Bonifacio07a}, and \citet{Hosford09}. To put constrainst on the primordial lithium abundance in the Universe, it is of fundamental importance to determine if and by how much the Li abundances increase with increasing metallicity on the Spite plateau. The behaviour of Li abundance with effective temperature is of equal importance, as it could lend support to the notion that atomic diffusion is acting in metal-poor stellar atmospheres. Lithium depletion through atomic diffusion has been suggested as a solution to the discrepancy between the Spite plateau abundance and the predicted value of the primordial lithium abundance \citep[see e.g.][]{Korn07,Lind09b}. To accurately infer both the mean lithium abundance in the halo and the abundance behaviour with metallicity and effective temperature, it is crucial to have a realistic description of the lithium line formation. Previous non-LTE analyses spanning a large stellar-parameter space \citep{Carlsson94,Pavlenko96,Takeda05} have shown that departures from LTE are generally small but significant at the required accuracy.
 
\citet{Barklem03b} applied quantum-mechanical calculations of cross sections for inelastic collisions with neutral hydrogen \citep{Belyaev03,Croft99} in non-LTE analysis of the Sun and two metal-poor stars. Their findings point to a negligible influence on the statistical equilibrium of lithium from collisional bound-bound transitions with hydrogen, but a significant influence from charge transfer reactions, specifically mutual neutralization and ion-pair production ($\rm Li^*+H \getsto Li^++H^-$). Other non-LTE investigations have relied on the classical Drawin recipes \citep{Drawin68}, as given by \citet{Steenbock84} and \citet{Lambert93}, or the free electron model of \citet{Kaulakys85}, for estimates of collisions with neutral hydrogen. None except Barklem et al.\ have included the influential charge transfer reaction.

We have extended the study by \citet{Barklem03b} to cover a large cool-star grid and calculate non-LTE abundance corrections for the lithium 670.7\,nm (2s-2p) and 610.3\,nm (2p-3d) lines.

\section{Setup}
We use the radiative-transfer code MULTI, version 2.3 \citep{Carlsson86,Carlsson92} to perform non-LTE calculations. The model atom used includes the same 20 energy levels for neutral lithium as described in \citet{Carlsson94}, plus the Li\,II ground state. The highest considered level in Li\,I has principle quantum number $n=9$. We have used TOPbase data \citep{Peach88} for energy levels, oscillator strengths, and photo-ionization cross-sections for levels with orbital quantum number $l\leq 3$. For the remaining levels, hydrogenic values are used. For the resonance line at 670.7\,nm, we adopt the oscillator strength $f=0.7468$, as calculated by \citet{Yan98}, and consider six hyperfine components in $^7$Li (neglecting $^6$Li), with measured wavelengths given by \citet{Sansonetti95}. For the subordinate line at 610.3\,nm, we adopt $f=0.6386$, also determined by Yan et al, and account for the three fine-structure components with wavelengths determined by \citet{Lindgard77}. For both lines, van-der-Waals-broadening parameters follow \citet{Anstee95} and \citet{Barklem97}. Stark broadening is unimportant in the late-type atmospheres of interest here and is therefore neglected. Table 1 lists wavelenghts, oscillator strengths, and broadening data adopted for the 670.7\,nm and 610.3\,nm lines.

Cross-sections for collisional excitation by electrons are taken from \citet{Park71} and collisional ionization by electrons from \citet{Seaton62b} as given by \citet{Allen76}.  We add rate coefficients for excitation and de-excitation from collisions with neutral hydrogen atoms according to \citet{Belyaev03} and \citet{Barklem03b}, as well as charge transfer reactions with neutral hydrogen and the negative hydrogen ion according to \citet{Croft99}. The number abundance of the negative hydrogen ion is calculated by assuming LTE. Ionization by hydrogen atom impact is not included. For low-lying states at low collision energies, it is expected to be negligible compared to charge transfer reactions and excitation \citep[see e.g.][]{Krstic09}. 

A grid of 1D, LTE, opacity-sampling, MARCS model atmospheres \citep{Gustafsson08} is used in the analysis. Non-LTE computations for lithium as a trace element are performed in the plane-parallel approximation, for models with $T_{\rm eff}=[4000,8000]$\,K, $\log{g}=[1.0,5.0]$, $\rm[Fe/H]=[0.0,-3.0]$, for lithium abundances in the range $A\rm(Li)=[-0.30,-4.20]$. The highest effective temperature is 5500\,K for models with $\log{g}=1.0$, 6500\,K for $\log{g}=2.0$, 7500\,K for $\log{g}=3.0$, and 8000\,K for $\log{g}\ge4.0$. For models with $\log{g}\ge3.0$, we adopted a microturbulence parameter $\xi_{\rm t}=1.0\rm\,km\,s^{-1}$ and $\xi_{\rm t}=2.0\rm\,km\,s^{-1}$, and for models with $\log{g}\le3.0$, we adopted $\xi_{\rm t}=2.0\rm\,km\,s^{-1}$ and $\xi_{\rm t}=5.0\rm\,km\,s^{-1}$. In total, 392 atmospheric models are included in the computations.   

In the statistical equilibrium calculations, the background opacities include continuum and line opacities provided by the MARCS model atmospheres. MULTI thus calculates a line-blanketed photo-ionizing radiation field while solving for the statistical equilibrium of lithium. Based on the LTE and non-LTE equivalent widths obtained for the stellar-parameter grid, we subsequently define a non-LTE correction for each abundance point as the difference between its LTE lithium abundance and the non-LTE abundance that corresponds to the same equivalent width. The limiting equivalent widths for which corrections are given are set to 0.01\,pm and 100\,pm. The equivalent width is obtained by numerical integration over the line profile, considering a spectral region that extends $\pm9\rm\,nm$ from the line centre. For lines weaker than 50\,pm the numerical precision is better than 0.01\,pm whereas the equivalent widths of the strongest lines are determined to within 0.1\,pm.

\begin{table}
      \caption{Wavelengths, oscillator strengths, and broadening data for the two considered lines.}
         \label{tab:data}
         \centering
         \begin{tabular}{llllll}
                \hline\hline
\multicolumn{6}{c}{$\rm 2s^2S-2p^2P^{o}$ }\\
\multicolumn{6}{c}{$^{(a)}\Gamma=3.690\times10^{7}$ \ \ \ $^{(b)}\sigma=346$ \ \ \  $^{(c)}\alpha=0.236$   }\\
\hline
  $\lambda$[nm]    &   $J_{\rm l}$  & $J_{\rm u}$&   $f$ & $F_{\rm l}$  & $F_{\rm u}$  \\
\hline
 670.79080     & 1/2            &1/2              & $1.037\times 10^{-2}$  &1             & 1     \\
 670.79066     & 1/2            &1/2              & $5.186\times 10^{-2}$  &1             & 2     \\
 670.79200     & 1/2            &1/2              & $3.112\times 10^{-2}$  &2             & 1     \\
 670.79187     & 1/2            &1/2              & $3.112\times 10^{-2}$  &2             & 2     \\
 670.77561     & 1/2            &3/2              & $1.245\times 10^{-1}$  &1             & 0,1,2 \\
 670.77682     & 1/2            &3/2              & $1.245\times 10^{-1}$  &2             & 1,2,3 \\    
             \hline
\\
\multicolumn{6}{c}{$\rm 2p^2P^{o}-3d^2D$ }\\
\multicolumn{6}{c}{$^{(a)}\Gamma=1.055\times10^{8}$ \ \ \ $^{(b)}\sigma=837$ \ \ \  $^{(c)}\alpha=0.274$   }\\
\hline
  $\lambda$[nm]    &   $J_{\rm l}$  & $J_{\rm u}$&    $f$  \\
\hline
610.3538     & 1/2             &3/2              & $6.386\times 10^{-1}$  &    &   \\ 
610.3664     & 3/2             &3/2              & $6.386\times 10^{-2}$  &    &   \\
610.3649     & 3/2             &5/2              & $5.747\times 10^{-1}$  &    &   \\
\hline
\\
\multicolumn{6}{l}{$^{(a)}$  $\Gamma$ $\rm[rad\,s^{-1}]$ is the natural broadening parameter.}\\
\multicolumn{6}{p{7cm}}{$^{(b)}$  $\sigma$ [a.u.\,] is the broadening cross-section for collisions with neutral hydrogen at relative velocity $v=10^4\rm m\,s^{-1}$. \citep{Anstee95}
}\\
\multicolumn{6}{p{7cm}}{$^{(c)}$  $\alpha$ is the velocity dependence of $\sigma$.}
         \end{tabular}
\end{table}

\section{Results}
The abundance corrections and LTE and non-LTE equivalent widths for each grid point are given in Table 2 for the 670.7\,nm line, and in Table 3 for the 610.3\,nm line. In the following we summarize the most important non-LTE effects and describe the size of the corrections over the cool-star grid. Because our results are in qualitative agreement with \citet{Carlsson94}, we refer the reader to this paper for a detailed explanation of the non-LTE line formation of lithium. 

At $A\rm(Li)<2$ for the coolest giants and $A\rm(Li)<3$ for the hottest giants, the dominant non-LTE effect is over-ionization of neutral lithium, driven mainly by the super-thermal radiation field at $\lambda\la349.5$\,nm ($J_\nu>B_\nu$, where $J_\nu$ is the mean intensity and $B_\nu$ the Planck function), corresponding to the photo-ionization edge for the first excited level. This leads to smaller number populations of the ground (2s) and first excited (2p) states compared to LTE and in higher values of the line source functions. Both the effect on the source functions and the loss of line-opacity weaken the lines in non-LTE, making the abundance corrections positive. The corrections range from being close to zero for the hottest giants to a maximum $+0.45$\,dex for cool, metal-rich giants for the 670.7\,nm line. For the 610.3\,nm line, the corrections are slightly smaller, $\leq+0.3$\,dex. 
 
In dwarfs, the $J_\nu-B_\nu$ excess is smaller in the 2p photo-ionization continuum and the radiation field is sub-thermal, $J_\nu<B_\nu$, in other influential continua, e.g. those corresponding to photo-ionization from 3p and 3d. In layers close to continuum optical depth unity, hot dwarfs have overpopulations in the ground state, which in combination with a decrease in the resonance-line source function strengthens the 670.7\,nm line. Abundance corrections are therefore negative or approximately zero for this line for dwarfs hotter than $6000\,$K.

At $A\rm(Li)>2$ for the coolest stars and $A\rm(Li)>3$ for the hottest stars, photon losses in the resonance line become apparent, driving recombination from Li\,II in both dwarfs and giants. This mainly affects the population of the ground state and to a lesser degree higher excited states. The resonance-line source function drops accordingly and the non-LTE corrections become less positive or more negative with increasing abundance over the whole grid. For metal-poor, extremely lithium-rich giants, the corrections can reach almost $-1.0$\,dex.

Since the number population of the first excited state and the source function of the subordinate line are not as severely affected by the photon losses in the resonance line, the 610.3\,nm abundance corrections for $A\rm(Li)<4.0$ are approximately constant for fixed stellar parameters. 

As described in \citet{Barklem03b}, including hydrogen collisions strengthens the collisional coupling between singly ionized lithium and neutral lithium, especially through the charge transfer reaction involving $\rm H$ and Li\,I in the 3s state and its inverse reaction. The number populations of low excited states increase when charge transfer is included, while the line source functions are barely affected. In turn, the abundance corrections become generally lower, i.e.\ less positive or more negative. Notice that the added collisions do not always have a thermalising effect, but can contribute to over-population of levels compared to LTE. The effect is illustrated in Fig.\,1, for a selected metal-poor and solar-metallicity dwarf and giant. As seen in Fig. 1, the change in $A\rm(Li)_{\rm non-LTE}-\it A\rm(Li)_{\rm LTE}$ when including charge transfer is $-0.12$\,dex for the metal-poor giant and $-0.06$\,dex for the metal-poor dwarf. 

While the charge transfer reaction has a significant influence on the statistical equilibrium of lithium, including bound-bound (and bound-free) transitions due to collisions with neutral hydrogen has no influence at all. However, this is not true if one relies on the classical Drawin recipe for estimates of the cross-section for those collisions. For comparison, Fig. 1 also shows the results obtained when neglecting charge transfer reactions but including excitation and ionization by neutral hydrogen according to Drawin's recipe, with the rate formula given by \citet{Lambert93} and with the scaling factor $S_{\rm H}=1.0$. Generally, the Drawin cross-sections are much higher than the quantum mechanical calculations by Belyaev \& Barklem. When adopting the higher classical rates, the ground state is more populated and the abundance corrections become lower, by $\sim-0.05$\,dex, compared to the results obtained with quantum mechanical rates. 

\begin{figure}
        \centering
                \includegraphics[angle=90,width=9.8cm]{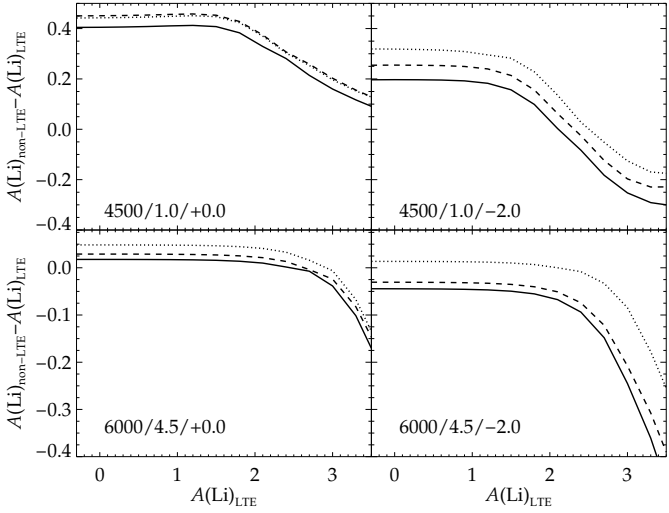}
          \caption{Non-LTE abundance corrections for 670.7\,nm as functions of LTE lithium abundance for the indicated stellar parameters in the lower left corner of each plot, to be read as $T_{\rm eff}/\log{g}/\rm[Fe/H]$. $Solid$: Including charge transfer reactions with hydrogen as well as bound-bound transitions due to collisions with neutral hydrogen. The cross-sections are calculated with quantum mechanics (Belyaev \& Barklem 2003; Croft et al.\ 1999). $Dotted$: Neglecting charge transfer reactions, but including bound-bound transitions, calculated with quantum mechanics. $Dashed$: Neglecting charge transfer reactions, but including hydrogen collisions for bound-bound and bound-free transitions. The cross sections are calculated with the classical Drawin recipe.}
        \label{fig:pic2}
\end{figure}
\begin{figure}
        \centering
                \includegraphics[angle=90,width=9.8cm]{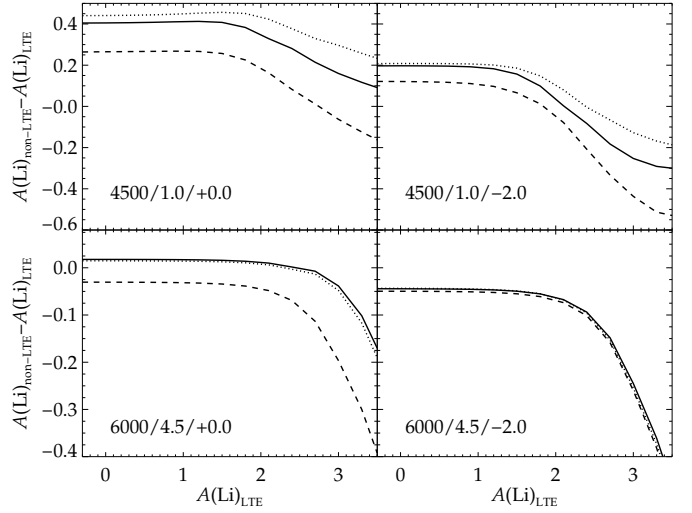}
          \caption{Panels as in Fig.\,1. $Solid$: Same as in Fig.\,1. MARCS model atmospheres from \citet{Gustafsson08} are used, with $\xi_{\rm t}=2\rm\,km\,s^{-1}$. $Dashed$: MARCS model atmospheres from \citet{Gustafsson75} are used, with $\xi_{\rm t}=2\rm\,km\,s^{-1}$. $Dotted$: MARCS model atmospheres from \citet{Gustafsson08} are used, with microturbulence parameter set to $\xi_{\rm t}=5\rm\,km\,s^{-1}$ in the two upper panels and $\xi_{\rm t}=1\rm\,km\,s^{-1}$ in the lower panels.}
        \label{fig:pic1}
\end{figure}

\begin{table}
\centering
      \caption{Non-LTE abundance corrections and equivalent widths for the Li\,I $\lambda=670.7$\,nm line. The whole table can be retrieved in machine-readable format from the CDS.}
      \begin{tabular}{p{0.6cm}p{0.6cm}p{0.6cm}p{0.7cm}p{0.9cm}p{0.6cm}p{0.7cm}p{0.7cm}}
                \hline\hline
$T_{\rm eff}$  & $\log{g}$& [Fe/H]      &$\xi_{\rm t}$  & $A\rm(Li)_{LTE}$  & $\Delta^{(a)}$ & $W_{\lambda\rm,LTE}$ & $W_{\lambda\rm,NLTE}$\\
 $\rm[K]$     &          &             & [km/s]         &        &          &   [pm]     &  [pm]\\
                \hline
4000 & 1.0 & -3.0 & 2.0 & -0.3 &  0.25&   2.24&   1.29\\
4000 & 1.0 & -3.0 & 2.0 &  0.0 &  0.25&   4.20&   2.49\\
4000 & 1.0 & -3.0 & 2.0 &  0.3 &  0.25&   7.48&   4.65\\
4000 & 1.0 & -3.0 & 2.0 &  0.6 &  0.24&  12.28&   8.28\\
4000 & 1.0 & -3.0 & 2.0 &  0.9 &  0.21&  18.05&  13.66\\
4000 & 1.0 & -3.0 & 2.0 &  1.2 &  0.14&  23.71&  20.40\\
...  &...  & ...  &...  &...   &...   &...    &...  \\
                \hline
\multicolumn{8}{l}{$^{(a)}$ $\Delta=A\rm(Li)_{non-LTE}-\it A\rm(Li)_{LTE}$}
      \end{tabular}
\end{table}

\begin{table}
\centering
      \caption{Non-LTE abundance corrections and equivalent widths for the Li\,I $\lambda=610.3$\,nm line. The whole table can be retrieved in machine-readable format from the CDS.}
      \begin{tabular}{p{0.6cm}p{0.6cm}p{0.6cm}p{0.7cm}p{0.9cm}p{0.6cm}p{0.7cm}p{0.7cm}}
                \hline\hline
$T_{\rm eff}$  & $\log{g}$& [Fe/H]      &$\xi_{\rm t}$       & $A\rm(Li)_{LTE}$  & $\Delta^{(a)}$ & $W_{\lambda\rm,LTE}$ & $W_{\lambda\rm,NLTE}$\\
 $\rm[K]$     &          &             & [km/s]         &        &          &   [pm]     &  [pm]\\
                \hline
4000 & 1.0 & -3.0 & 2.0 &  0.0 &  0.17&   0.02&   0.02\\
4000 & 1.0 & -3.0 & 2.0 &  0.3 &  0.17&   0.05&   0.03\\
4000 & 1.0 & -3.0 & 2.0 &  0.6 &  0.17&   0.09&   0.06\\
4000 & 1.0 & -3.0 & 2.0 &  0.9 &  0.17&   0.19&   0.13\\
4000 & 1.0 & -3.0 & 2.0 &  1.2 &  0.17&   0.37&   0.25\\
4000 & 1.0 & -3.0 & 2.0 &  1.5 &  0.17&   0.73&   0.49\\
...  &...  & ...  &...  &...   &...   &...    &...  \\
                \hline
\multicolumn{8}{l}{$^{(a)}$ $\Delta=A\rm(Li)_{non-LTE}-\it A\rm(Li)_{LTE}$}
      \end{tabular}
\end{table}

\begin{figure}
        \centering
                \includegraphics[angle=90,width=9.8cm]{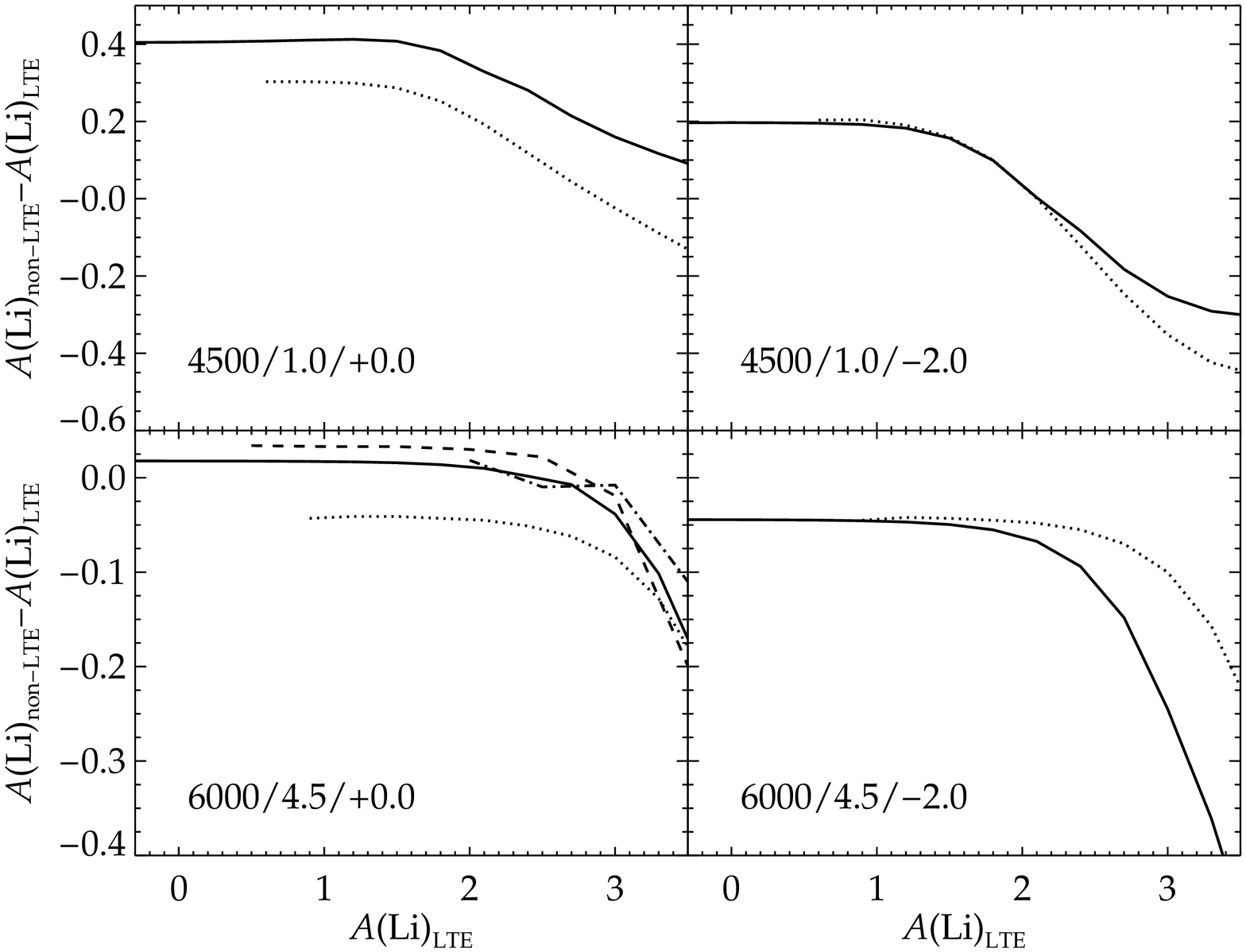}
          \caption{Panels as in Fig.\,1. $Solid$: Same as in Fig.\,1 and 2. $Dotted$: \citet{Carlsson94} corrections. $Dashed$: \citet{Takeda05} corrections. $Dash-dotted$: \citet{Pavlenko96} corrections.}
        \label{fig:pic3}
\end{figure}

\section{Discussion}
We have compared our results for the lithium resonance line to \citet{Carlsson94} abundance corrections for dwarfs and giants with $\rm[Fe/H]=[-3.0,0.0]$ to \citet{Pavlenko96} results for dwarfs and subgiants with solar metallicity and to \citet{Takeda05} corrections for dwarfs and subgiants with $\rm[Fe/H]=[0.0,-1.0]$ (see Fig.\,3). To aid in the comparison we converted the LTE and non-LTE equivalent widths listed in Table 1 in \citet{Pavlenko96} to abundance corrections. The largest difference is 0.12\,dex between our results and those of \citeauthor{Pavlenko96} and 0.15\,dex between our results and those of \citeauthor{Takeda05}, but generally the values agree quite well. The \citet{Carlsson94} values agree with ours to within 0.15\,dex for dwarfs and 0.20\,dex for giants. Such differences are reasonable considering differences in the model atom and in the atmospheric models. \citeauthor{Carlsson94} use \citet{Gustafsson75} MARCS models and \citet{Takeda05} and \citet{Pavlenko96} use \citet{Kurucz93} ATLAS9 models. The effects on the non-LTE abundance corrections when using MARCS models from \citet{Gustafsson75} and \citet{Gustafsson08} are illustrated in Fig\,2. Especially for the two giants, the choice of model atmosphere is important for the corrections. The newer models include more line opacity, causing a steeper temperature gradient in the upper part of the photosphere, which increases the ultraviolet $J_\nu-B_\nu$ excess and leads to more over-ionization. The use of newer model atmospheres thus leads to more positive or less negative abundance corrections and partly cancels with the effect of including hydrogen collisions (see Sect.\ 3)

Varying the microturbulence between $1\rm\,km\,s^{-1}$ and $2\rm\,km\,s^{-1}$ barely affects the abundance corrections (lower panels in Fig.\,2), especially for low lithium abundances. In a giant star, the non-LTE abundance corrections become systematically higher when adopting a microturbulence of $5\rm\,km\,s^{-1}$ instead of $2\rm\,km\,s^{-1}$, the differences being especially significant when the line is strong (upper panels Fig.\,2). However, the parameter still has some significance at low lithium abundances, when the formation of the line itself is unaffected by the microturbulence. This is because the choice of microturbulence influences the amount of line opacity included in the computations of the model atmosphere and consequently also the atmospheric temperature gradient, which in turn partly drives the departures from LTE.

Our results are valid within the assumptions of 1D model atmospheres in LTE and hydrostatic equilibrium. Non-LTE calculations for lithium in 3D, hydrodynamical model atmospheres have to this date been performed only for a few types of stars \citep{Asplund03}, and we plan to extend our work to 3D in the future.

\acknowledgements{We are grateful to Mats Carlsson for his helpful support and comments on this work. K.\,L.\ thanks Jon Sundqvist for careful reading of the manuscript. P.\,S.\,B.\ is a Royal Swedish Academy of Sciences Research Fellow supported by a grant from the Knut and Alice Wallenberg Foundation. P.\,S.\,B.\ also acknowledges additional support from the Royal Swedish Academy of Sciences and the Swedish Research Council.}

\bibliographystyle{aa}

\end{document}